\documentclass[12pt, draftclsnofoot, onecolumn]{IEEEtran}
\usepackage{graphicx,amssymb,amsmath}
\usepackage{subfigure}
\usepackage{multicol}
\usepackage[noadjust]{cite}
\usepackage{setspace}
\usepackage{stfloats}
\usepackage{midfloat}
\usepackage[normal]{threeparttable}
\usepackage{amsthm}
\usepackage{cuted}
\usepackage{cases,subeqnarray}
\usepackage{bm,multirow,bigstrut}
\usepackage{textcomp}
\usepackage{latexsym,bm}
\usepackage{booktabs,changebar}
\usepackage{xcolor}
\usepackage{mathtools}
\usepackage{dsfont}
\usepackage{extarrows}
\usepackage{caption2}
\theoremstyle{plain}

\theoremstyle{plain}

\IEEEoverridecommandlockouts
\begin{document}
\title{{On Low-Resolution ADCs in Practical 5G Millimeter-Wave Massive MIMO Systems}}
\author{Jiayi~Zhang,
        Linglong~Dai,
        Xu~Li,
        Ying~Liu,
        and~Lajos Hanzo
\thanks{{\copyright IEEE Commun. Mag. 2018}. This work was supported in part by the National Natural Science Foundation of China (Grant No. 61601020), the Beijing Natural Science Foundation (Grant Nos. 4182049 and L171005), the open research fund of National Mobile Communications Research Laboratory, Southeast University (Grant No. 2018D04), the Fundamental Research Funds for the Central Universities (Grant Nos. 2016RC013, 2017JBM319, and 2016JBZ003), and the Royal Academy of Engineering through the UK-China Industry Academia Partnership Programme Scheme (Grant No. UK-CIAPP$\backslash$49). (Corresponding author: Linglong Dai and Lajos Hanzo.)}
\thanks{J. Zhang is with the School of Electronics and Information Engineering, Beijing Jiaotong University, Beijing 100044, P. R. China. He is also with National Mobile Communications Research Laboratory, Southeast University, Nanjing 210096, P. R. China.}
\thanks{L. Dai is with Department of Electronic Engineering as well as Tsinghua National Laboratory of Information Science and Technology (TNList), Tsinghua University, Beijing 100084, P. R. China.}
\thanks{X. Li and Y. Liu are with the School of Electronics and Information Engineering, Beijing Jiaotong University, Beijing 100044, P. R. China.}
\thanks{L. Hanzo is with the School of ECS, University of Southampton, UK.}
}

\maketitle
\begin{abstract}
Nowadays, millimeter-wave (mmWave) massive multiple-input multiple-output (MIMO) systems is a favorable candidate for the fifth generation (5G) cellular systems. However, a key challenge is the high power consumption imposed by its numerous radio frequency (RF) chains, which may be mitigated by opting for low-resolution analog-to-digital converters (ADCs), whilst tolerating a moderate performance loss. In this article, we discuss several important issues based on the most recent research on mmWave massive MIMO systems relying on low-resolution ADCs. We discuss the key transceiver design challenges including channel estimation, signal detector, channel information feedback and transmit precoding. Furthermore, we introduce a mixed-ADC architecture as an alternative technique of improving the overall system performance. Finally, the associated challenges and potential implementations of the practical 5G mmWave massive MIMO system {with ADC quantizers} are discussed.
\end{abstract}

\IEEEpeerreviewmaketitle
\section{Introduction}
Both millimeter-wave (mmWave) and massive multiple-input multiple-output (MIMO) systems have emerged as attractive technical enablers of achieving a 1000-fold system throughput improvement for the fifth generation (5G) cellular systems \cite{boccardi2014five}. MmWave communications utilize the spectrum spreading from 30 to 300 GHz, where the spectrum is less crowded. The major benefit of using mmWave spectrum is that much larger bandwidths are available (e.g., 1 GHz or more) than that in the operational wireless systems relying on no more than 6 MHz. However, the key challenge of mmWave communications is the much higher propagation attenuation of mmWave signals than that of conventional low frequency signals (e.g., sub-6 GHz). Fortunately, massive MIMO systems are capable of providing high beamforming gains to compensate for the severe signal attenuation of mmWave, which is realized by directional transmissions relying on high-dimensional antenna arrays. As an additional benefit, more antennas can be fitted into the same array space, because the wavelength is shorter than that of the conventional signal frequencies below 6 GHz. Thanks to the large bandwidth of mmWave carrier frequencies and the high degree-of-freedom of large-scale antennas, mmWave massive MIMO systems are able to provide very high-speed data rate in cellular networks.


{Yet, the potentially high power consumption of mmWave massive MIMO systems may prevent their practical implementation for 5G. Future mmWave massive MIMO aided base stations (BSs) may use hundreds of antennas (either co-located or geographically distributed), where each antenna is connected to a dedicated radio-frequency (RF) chain. Typically, in Fig. 1, every receive RF chain consists of two analog-to-digital converters (ADCs), demodulator, down-converter, low noise amplifier (LNA), mixers, automatic gain control (AGC), variable gain amplifier (VGA), and some filters. Similar circuitry is employed at the transmit side. Although the advances in mmWave chip fabrication have significantly reduced the cost of electronics, the power consumption of ADCs still dominate the total power consumption of the whole RF chain.Firstly, {the power consumption of ADCs is linearly proportional to the sampling rate}, which is high due to the huge bandwidth of mmWave signals. Secondly, in a $b$-bit ADC relying on the typical flash architecture, the power consumption grows exponentially with the ADC resolution $b$. Typically, the power consumption of today's commercial high-speed ($\geq 20$ GSample/s), high-resolution (namely that 8-12 bits) ADCs is around 500 mW. For a mmWave massive MIMO system with 256 RF chains and 512 ADCs, the total power consumption of ADCs will become as high as 256 W, which is potentially unaffordable for the practical mmWave massive MIMO system. {Furthermore, the power consumption of both front-end and baseband circuitry is also high when high-resolution ADCs are used.}}

\begin{figure}[t]
\centering
\includegraphics[scale=1]{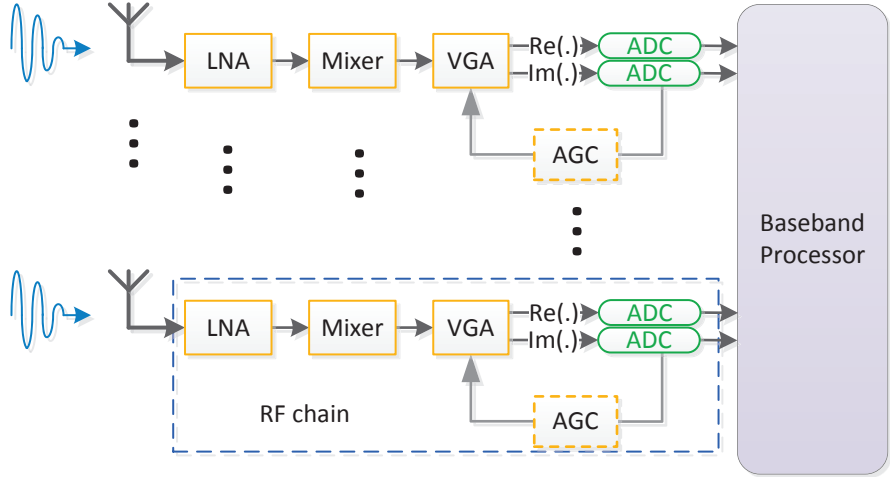}
\caption{MmWave massive MIMO receiver architecture with low-resolution ADCs. Each receiver antenna is connected with two low-resolution ADCs, where the quadrature and in-phase elements of the received signal at every antenna are quantized separately. The function of a VGA and an AGC before the ADC is to assure that the power of analog signals is within a right range. The output of quantization is used for digital baseband processing.
\label{fig:low_resolution_ADC}}
\end{figure}

{To alleviate this predicament, on one hand, we can employ several high-resolution, low-speed sub-ADCs operating in parallel. Yet, this ADC structure may impose error floors on the system's performance due to the mismatch among the sub-ADCs. On the other hand, we can opt for high-speed but low-resolution (namely that 1-3 bits) ADCs {to decrease both the power consumption and the hardware cost.}} The architecture of mmWave massive MIMO systems based on low-resolution ADCs is illustrated in Fig. \ref{fig:low_resolution_ADC}, {where every RF chain is connected to two low-resolution ADCs rather than high-resolution ADCs.} Although the use of low-resolution ADCs can significantly decrease the circuit power consumption and relax the requirement of accurate baseband circuitry, further challenges are imposed on the signal processing algorithms. For example, {the rough nonlinear distortion} inflicted by low-resolution ADCs may render the signal processing algorithms developed for high-resolution ADCs suboptimal. Moreover, compared to conventional low frequency channels, the mmWave channel exhibits spatial/angular domain sparsity resulting from {the employment of densely-spaced array elements} and large bandwidths. Only a limited number of dominant multipath components (e.g., 2-5 paths) exhibit the typical mmWave multipath channel. {Since the channel response matrix exhibits a low row rank and a big condition number, low-complexity precoding of channel inversion may not be optimum.}

\begin{figure}[htbp]
\centering
\includegraphics[scale=0.8]{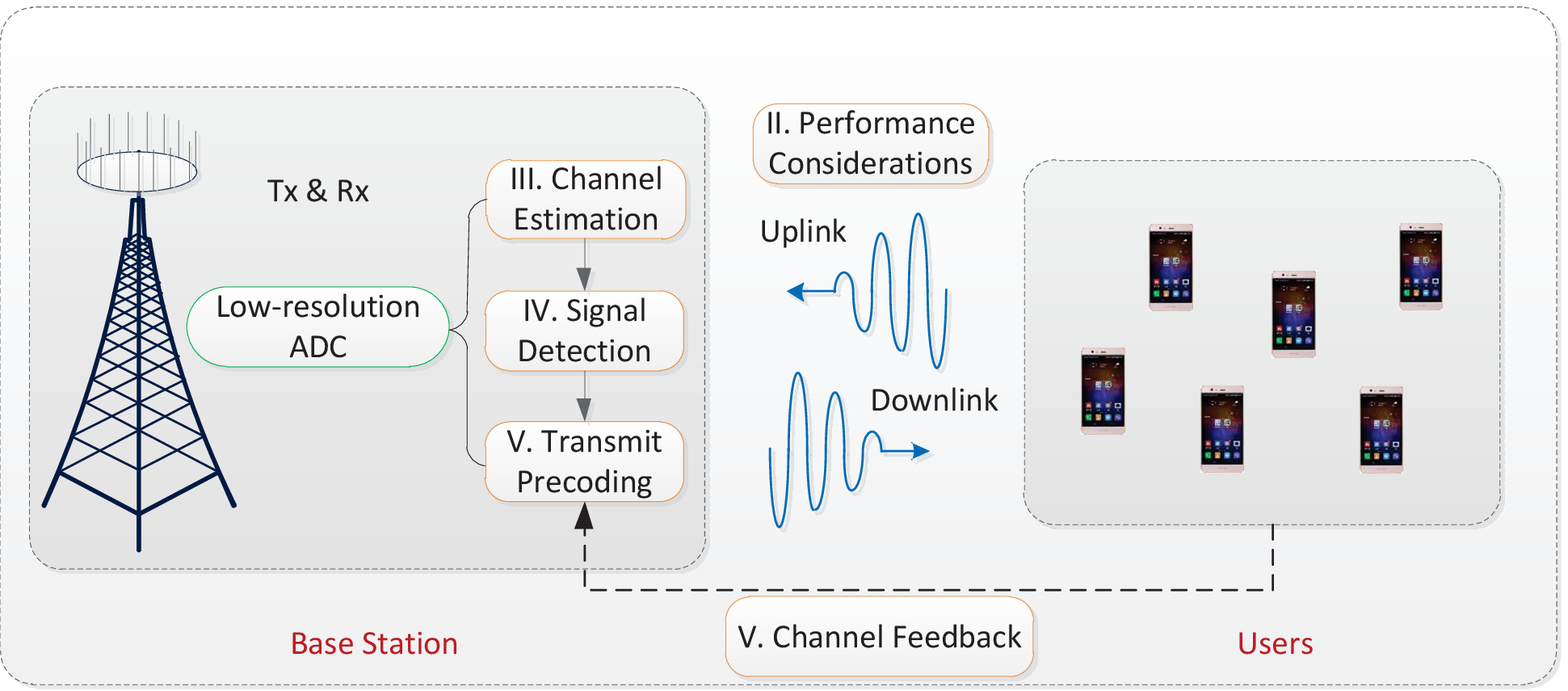}
\caption{Key challenges and potential future research directions for practical mmWave massive MIMO systems relying on low-resolution ADCs from a signal processing perspective. More specifically, the performance decrease due to low-resolution ADCs is analyzed in Section II. We study how to utilize the spatial/angular sparsity of the mmWave massive MIMO channel to design efficient channel estimation algorithms in Section III. Novel signal detection schemes conceived for counteracting the deleterious effects of low-resolution ADCs are studied in Section IV. Finally, Section V proposes new codebook designs for CSI feedback and the class of CSI-based transmit precoding algorithms.}
\label{fig:Structure_signal}
\end{figure}

{As shown in Fig. \ref{fig:Structure_signal}, in this article we consider practical mmWave massive MIMO systems relying on low-resolution ADCs from a signal processing perspective, highlighting the key challenges and a range of promising future research directions. Specifically, in Section II we first investigate the performance of mmWave massive MIMO systems using low-resolution ADCs. Then in Section III, we study how we can exploit the angular and spatial sparsity of the mmWave massive MIMO channel for designing efficient channel estimation algorithms. Given the focus of CSI, Section IV is on novel signal detection schemes conceived for counteracting the deleterious effects of low-resolution ADCs. Additionally, new codebook designs are proposed for CSI feedback and the class of CSI-based transmit precoding algorithms is elaborated on in Section V. Finally, sophisticated techniques of mitigating the strong nonlinearity distortion imposed by coarse ADC quantization are proposed and future research directions are discussed in Section VI.}



\section{{Performance Considerations}}
{The performance of quantized mmWave massive MIMO systems with low-resolution ADCs is lower than that of the idealized one operating without quantization.} It is widely recognized in information theory that the optimal distribution of the transmitted signal has to be designed to achieve the capacity of continuous memoryless channel. However, the channel capacity of realistic mmWave massive MIMO systems using low-resolution ADCs is approached by the discrete input distribution \cite{mo2015capacity}. In order to approach the capacity, the transmitter should have knowledge of the channel state information (CSI) to design the transmitted constellation. It has been shown in \cite{singh2009limits} that binary antipodal signaling is optimal in the context of single-input single-output channels for one-bit ADCs. {High-order constellations, for example 16-QAM, are able to be supported} by spatial oversampling of one-bit ADC based massive MIMO systems \cite{jacobsson2016massive}. However, the optimal signaling alphabet is still unknown for the practical mmWave massive MIMO channel. {One can design the input alphabet by using a computationally efficient approach based on convex optimization.}

\begin{figure}
  \centering
  \subfigure[{Average spectral efficiency of massive MIMO systems relying on low-resolution ADCs over Rician fading channels. The average transmit SNR of each user is 10 dB, the number of users is 10, and the Rician-$K$ factor is 100.}]{
    \label{fig:Ana_MRC} 
    \includegraphics[width=2.8in]{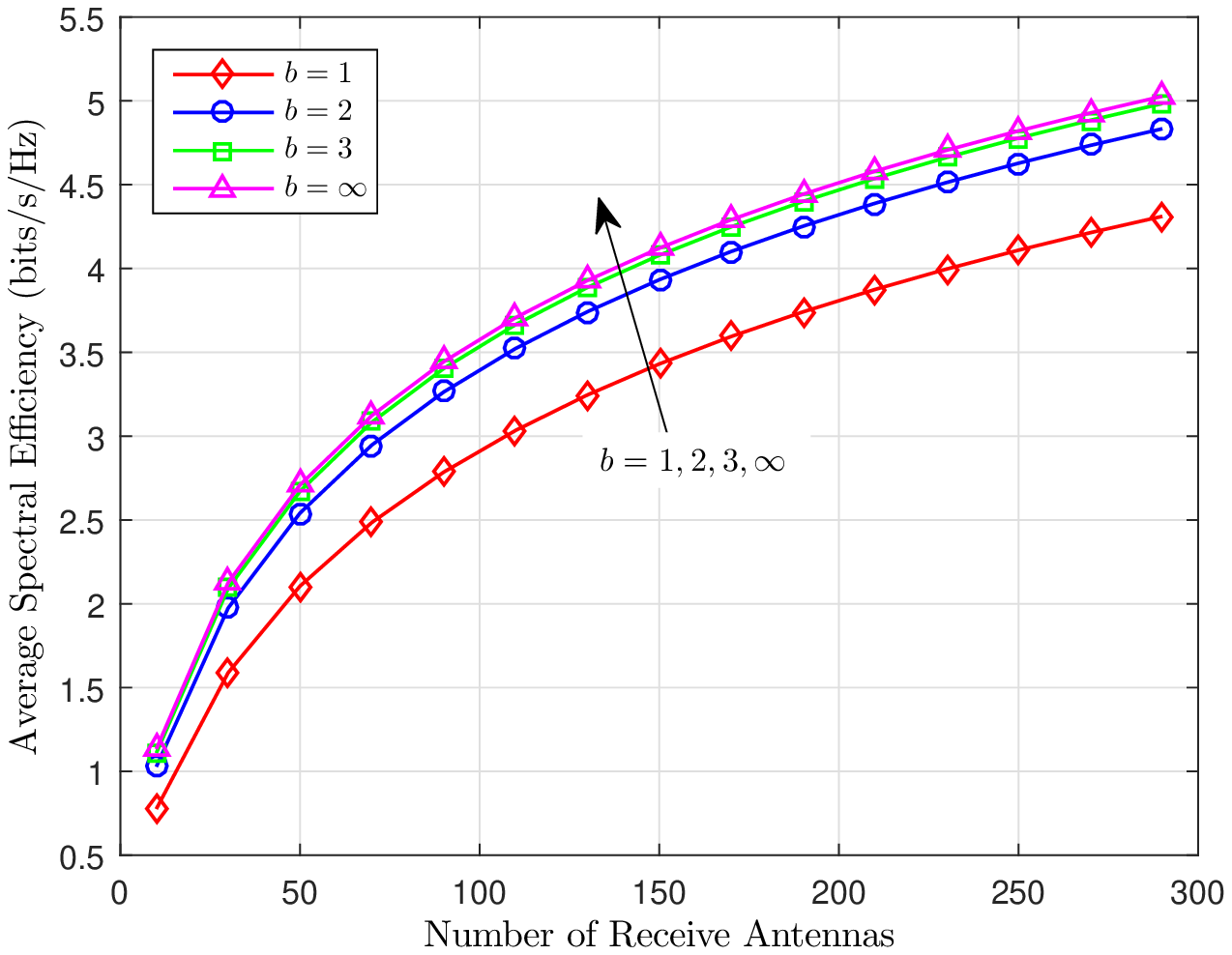}}
    \hspace{0.3in}
  \subfigure[{Average spectral efficiency of massive MIMO systems relying on mixed ADCs over Rician fading channels. The average transmit SNR of each user is 10 dB, the number of users is 10, and the Rician-$K$ factor is 100.}]{
    \label{fig:Ana_MRC_mix} 
    \includegraphics[width=2.8in]{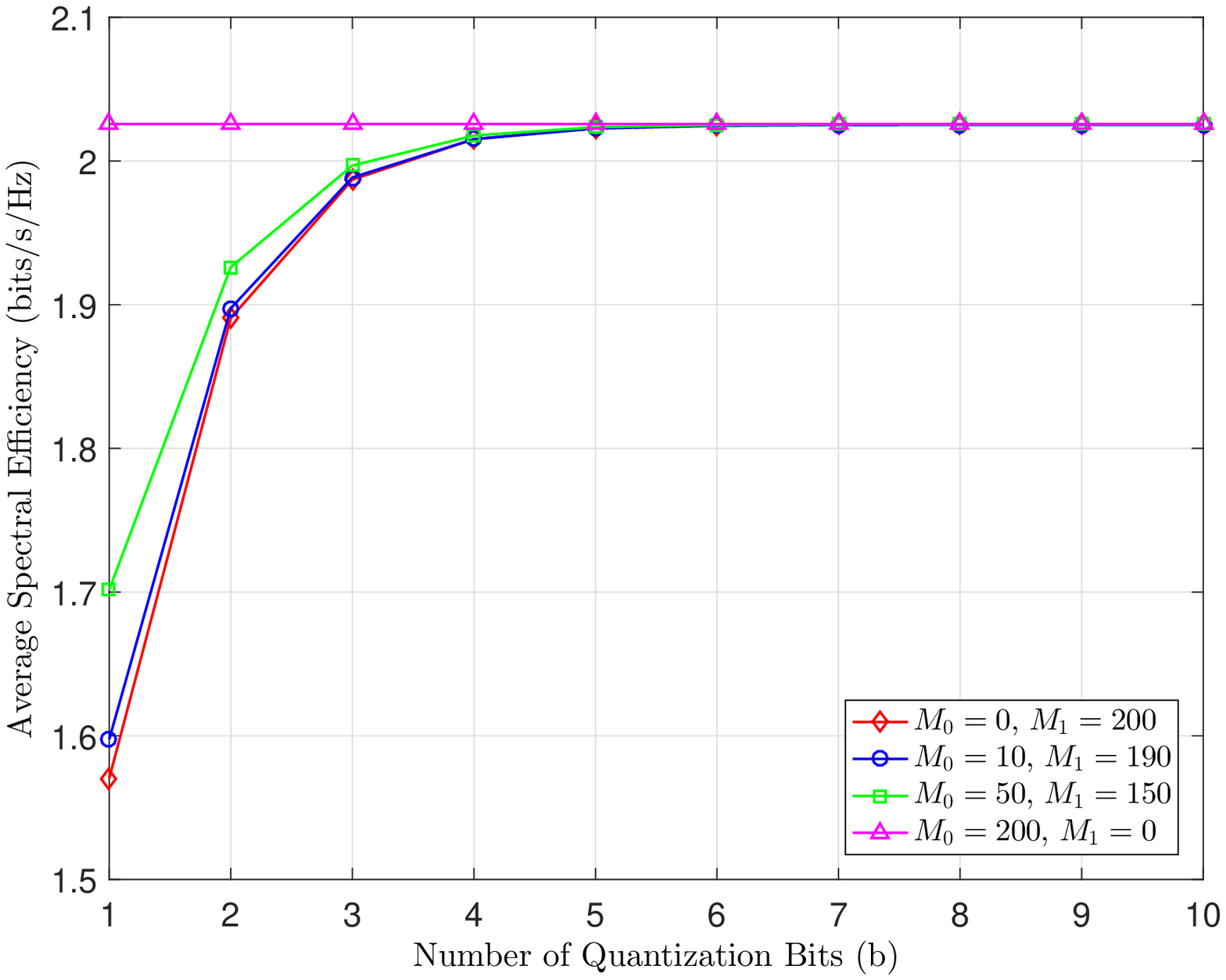}}
  \caption{Average spectral efficiency of quantized massive MIMO systems.}
  \label{fig:subfig} 
\end{figure}

{The widely used additive quantization noise model (AQNM) can characterize the quantization noise} \cite{zhang2016spectral,mo2015capacity,bai2015energy}. Under the assumption of Gaussian distributed signaling alphabets, a AQNM based lower capacity bound can be derived to provide valuable insights into the performance of quantized mmWave massive MIMO systems. In \cite{zhang2016spectral}, we have obtained closed-form expressions for the attainable spectral efficiency by using the AQNM. Fig. \ref{fig:Ana_MRC} investigates the effect of {ADC quantizers} on the spectral efficiency of massive MIMO systems communicating over Rician fading channels. We {exploit} in Fig. \ref{fig:Ana_MRC} that the spectral efficiency of 3-bit ADCs is close to the idealized infinite resolution case of $b=\infty$, which demonstrates that the performance loss introduced by 3-bit ADCs is modest. Moreover, Fig. \ref{fig:Ana_MRC} indicates that {using more antennas can compensate} the spectral efficiency loss imposed by low-resolution ADCs. {This implies that massive MIMO systems can operate with ADC quantizers by employing more antennas. In Fig. \ref{fig:Ana_MRC_mix}, we show the huge performance gain of massive MIMO systems relying on mixed-ADCs, where $M_0$ high-resolution ADCs and $M_1$ low-resolution are deployed. Due to the little performance loss, it is compelling to employ the mixed-ADC architecture.}
%

{Most of the recent contributions on the performance analysis applied the AQNM, which is sufficiently accurate at low SNRs. However, at high SNRs, the AQNM is {an inaccurate} method due to three reasons. Firstly, {the input signal obey the continuous Gaussian distribution. Secondly, the quantization noise is also assumed to follow Gaussian distributed.} Thirdly, the Max-Loyd quantizer of the ADC minimizing the MSE is not necessarily optimal in terms of maximizing the system's capacity. The exact channel capacity of massive MIMO systems {with ADC quantizers} is still unknown, since the input distribution is unknown and so are appropriate quantizer thresholds}

For example, using one-bit quantization, the lower AQNM bound is only $73\%$ of the ideal-ADC-based capacity \cite{mo2015capacity}. From the perspective of information theory, we conclude that the system's performance remains adequate by using a large-scale antenna array associated with low-resolution ADCs in {low SNRs}. By contrast, in the high-SNR regime, adopting some RF chains using high-resolution ADCs shows great promise in terms of improving the system performance. 

{In order to} maximize the energy efficiency, the tradeoff between the ADC resolution and the power consumption of ADCs should be carefully investigated. Bai and Nossek demonstrated in \cite{bai2015energy} that a substantial energy efficiency improvement can be achieved by using less quantization bits at low SNRs. Moreover, {the optimal number of quantization bits is related on the average SNR} at the receiver. {As shown in \cite{bai2015energy}, the tradeoff between the ADC resolution and the power consumption of ADCs should be carefully investigated in order to maximize the energy efficiency.}

{A compelling spectral and energy efficiency trade-off may be achieved with the aid of {ADC quantizers} at low SNRs.} {Nonetheless}, improving the performance of mmWave massive MIMO systems {with ADC quantizers} still constitutes a research challenge. Although previous contributions have investigated the asymptotic capacity of quantized MIMO systems both at low and at high SNRs, further investigations are required for determining the exact capacity of quantized mmWave MIMO systems. Furthermore, it is an often-used idealized simplifying assumption of prior contributions that perfect CSI is {obtainable} at the receiver, which is unrealistic for quantized receiver processing. An {significant} direction for future work is to analyze the effect of quantising the received pilots on the performance. Another major challenge is to find the optimal thresholds of the ADC quantizer for maximizing both the energy efficiency and the spectral efficiency. Finally, the optimal signalling alphabet distribution of low-resolution ADCs using more than one bits remains an active area of research, especially for the mmWave channel. {Since the wavelength of mmWave carriers is very short compared to the typical size of objects in the propagation environment, the {performance investigation} of massive MIMO systems with {ADC quantizers} should consider the unique channel characteristics of mmWave scenarios.}

%
%

\section{{Channel Estimation}}\label{se:channel_estimation}
As discussed above, accurate knowledge of the CSI is essential for realizing the {promising} gain of mmWave massive MIMO systems, which is usually acquired by using pilot symbols. In the uplink, the users need to send a training pilot to the BS. Then, the BS invokes channel estimators for acquiring the CSI, and performs symbol detection based on the received signal. However, the severe nonlinearity of low-resolution ADCs degrades the performance of conventional channel estimation techniques. {Many channel coefficients are needed to be trained with direct employment of conventional channel estimation techniques}, which means that the training sequence should be long enough for ensuring a reliable CSI estimation. For example, it has been shown in \cite{mo2014channel} that upon using a long training sequence (e.g., 50 times more symbols than the {amount of users), MIMO systems with one-bit ADCs} can approach the performance of the unquantized scenario. Therefore, more efficient channel estimation techniques relying on short training sequence should be developed for mmWave massive MIMO systems with {ADC quantizers}.

Motivated by these considerations, previous work has proposed several channel estimation algorithms. Specifically, {the expectation-maximization (EM) algorithm repeatedly computes the minimum mean square estimate of the quantized received signals until it converges. However, EM has a high complexity, {because every} EM iteration requires computing a matrix inverse and a large number of iterations are required for achieving convergence \cite{mezghani2010multiple}.} The gap w.r.t. the unquatized case is relatively small in terms of the mean squared error (MSE) performance, especially at low SNRs. However, {the initialization of the EM algorithm is significant and may approach a local optimum in high SNRs}. Considering the sparsity of the mmWave channel impulse response (CIR) in the angular domain, a computationally efficient algorithm {mentioned as} generalized approximate message passing (GAMP) has been {presented} in \cite{mo2014channel}. Explicitly, the GAMP algorithm {changes} the vector-valued estimation problem into several scalar problems and {a superior performance than the EM algorithm can be obtained in the low and medium SNRs}. Moreover, we can convert the classic maximum likelihood (ML) estimation into a convex optimization problem, provided that the constraint on the legitimate transmitted symbols is relaxed. The ML channel estimator recently proposed in \cite{choi2015near} is {utilized off-the-shelf convex optimization approaches} and can efficiently estimate {not only the norm of the channel but also the direction}. The MSE performance of the ML estimator is better than that of the EM algorithm and of other linear estimation methods at high SNRs, while its performance is limited in low SNRs. However, the training overhead of conventional channel estimators still remains excessive for practical mmWave massive MIMO systems relying on low-resolution ADCs.

Based on the aforementioned observations, the challenge of obtaining acceptable CSI in quantized massive MIMO systems is that of requiring an the excessively long training sequence. One possible solution is to employ joint channel-and-data (JCD) estimation or decision-directed channel estimation (DDCE), which exploits the reliably detected payload data to act as pilot sequences for assisting in channel estimation. {In contrast to the pilot based channel estimation schemes, the optimal Bayesian JCD algorithm jointly estimates the channel matrix and data symbols at the same time. The Bayesian inference framework can be used for achieving the best MSE estimate.} As a benefit, both DDCE and JCD estimation requires a relatively short training sequence for achieving comparable performance to that of the idealized unquantized case. Moreover, a Bayes-optimal JCD estimator was proposed in \cite{wen2015bayes} to achieve the best possible joint channel response and payload data estimation performance. However, it remains an open challenge to {obtain} the performance of the perfect CSI case by using the Bayes-optimal JCD estimator. Furthermore, hundreds of iterations are required to converge at low SNRs, hence the computational complexity imposed may become excessive. Thus, it is not practical to employ the Bayes-optimal JCD algorithm in a commercial mmWave massive MIMO system. A possible option in this direction is to adopt reduced-complexity suboptimal methods. The JCD estimator based on the suboptimal least-squares (LS) algorithm combined with zero-forcing (ZF) or maximal ratio combining (MRC) receivers is capable of supporting both multi-user operation and high-order modulation \cite{jacobsson2016massive}.

{As for channel estimation, the popular channel estimators are concluded in Table I. The common problem of these estimators is that they require substantial pilot overheads. Furthermore, these channel estimation algorithms usually ignore the mmWave channels' characteristics, namely that there are fewer channel paths than the number of antennas.} Some contributions on channel estimation only consider the case of one-bit ADC quantization \cite{mezghani2010multiple,mo2014channel,choi2015near}. {In case of a high ADC resolution, the complexity of channel estimation can be reduced as a result of having more accurate quantization. For example, it was suggested in \cite{wen2015bayes} that the complexity of channel estimation will reduce half by using 3-bit ADCs instead of 1-bit ADCs.} {An significant direction for future study} is to exploit the benefit of using low-resolution ADCs combined with high-resolution channel estimation. For example, the mixed-ADC receiver of \cite{liang2015mixed} partially employs high-resolution ADCs to {conduct} channel estimation. The channel estimation error of each antenna approximately follows the Gaussian distribution, which helps reduce the estimation bias. Furthermore, most of the prior contributions neglected the sparse structure of the mmWave CIR. {More sophisticated algorithms should be designed for exploiting the {sparse property of the mmWave channel} experienced both in the temporal- and in the spatial-domains for reducing the implementation complexity. Generally speaking, the channel estimation problem of low-resolution quantized MIMO systems operating in mmWave channels can be formulated as {a recovery problem of sparse signal}, which could be efficiently solved by the powerful technique of compressed sensing.} Therefore, efficient compressed sensing techniques can be invoked at the receiver. Finally, since the requirement of long training sequences is a key obstacle of accurate channel estimation, greedy algorithms can be used for exploiting the mmWave channel statistics for circumventing this challenging problem.

Due to the long training sequence, deriving acceptable channel information for mmWave massive MIMO systems relying on low-resolution ADC may become infeasible. {Instead of coherent detection approaches, we may resort to low-complexity noncoherent detection approach \cite{hanzo2011near} relying on the concept of autocorrelation-based detection, the receiver may relax its dependence on channel information. Low-complexity noncoherent techniques are favorable for mmWave massive MIMO systems relying on low-resolution ADCs. {In order to further enhance the performance of system}, more efficient noncoherent detection schemes relying on phase-decoded multiple-symbol detection are favorable.}

\begin{table}[!t]
\renewcommand{\thetable}{\Roman{table}}
\caption{Comparison of Channel Estimation Algorithms for Quantized MIMO Systems}
\label{table}
\centering
\begin{tabular}{lrrrrr}
\toprule[2pt]
  Algorithms & MSE & Complexity     & Training length  & ADC    & Channel model \\
\midrule[1pt]
  EM \cite{mezghani2010multiple}    &  bad      &   low &   200 & one-bit & MIMO  \\
  LS \cite{jacobsson2016massive}     &  fair      &   low &   100 & multi-bit & massive MIMO \\
  GAMP \cite{mo2014channel}     &  good       &   medium &   64-256 & one-bit & mmWave MIMO  \\
  ML \cite{choi2015near}     &  good      &   high &   50-150 & one-bit & massive MIMO  \\
  JCD \cite{wen2015bayes}     &  very good      &   too high &   4-32 & multi-bit & massive MIMO  \\
 \bottomrule[2pt]
\end{tabular}
\end{table}

\section{{Signal Detection}}

Given the CSI, the received signals can be coherently detected at the receiver. Classical signal detectors, such as the MRC, ZF, LS, minimum mean square error (MMSE), ML and the message passing detector have been widely used for idealized unquantized MIMO systems. However, the performance of these classic receivers erodes in the face of low-resolution ADCs. Furthermore, there is no straightforward technique of extending the results based on frequency-flat channels to wideband mmWave channels. It has been shown in \cite{jacobsson2016massive} that the performance of MRC and ZF detectors relying on low-resolution ADCs suffers from substantial performance degradations at high SNRs. {Furthermore, the computational complexity of most signal detection techniques may still be excessive for practical mmWave massive MIMO systems.} Finally, these optimal detectors relying on high-resolution ADCs become suboptimal for low-resolution ADCs. Therefore, efficient new signal detectors have to be constructed for counteracting the deleterious effects of low-resolution ADCs and to exploit the CIR's sparsity. In \cite{wang2015multiuser}, a multiuser message passing based detector has been proposed for massive spatial modulation (SM) based MIMO systems, where only a single RF chain is {needed}, which has substantial benefits. Hence its performance is superior to that of other linear detectors \cite{wang2015multiuser}. Another important advantage is its low computational complexity especially in the context of massive MIMO systems. However, since it was developed for the special case of SM schemes, it may not be directly {proper} for general massive MIMO systems {with ADC quantizers}. Moreover, it has not considered the sparsity of the mmWave CIRs.

\begin{figure}[thbp]
\centering
\includegraphics[scale=0.8]{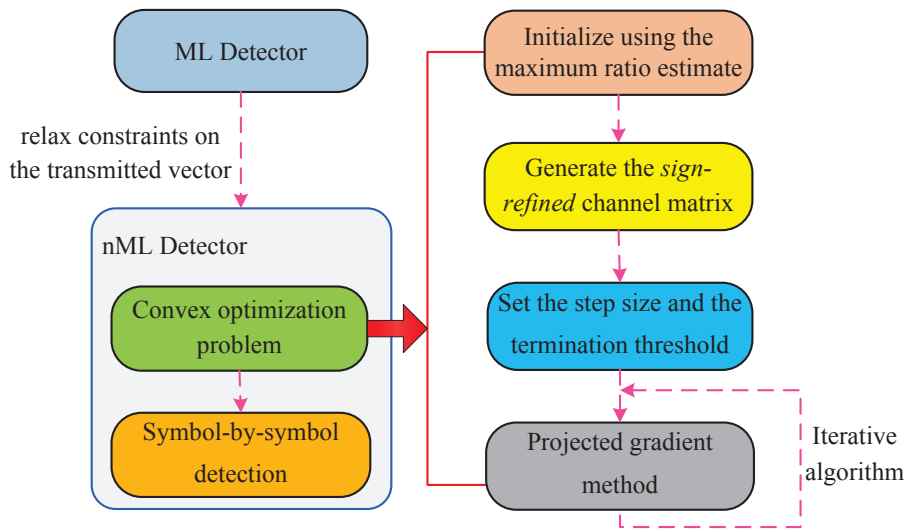}
\caption{The nML detector conceived for massive MIMO systems using one-bit ADCs.
\label{fig:ADC_signal_detector}}
\end{figure}

For {one-bit ADCs} massive MIMO systems, a near-ML (nML) detector was proposed in \cite{choi2015near}. As illustrated in Fig. \ref{fig:ADC_signal_detector}, the nML detector is based on well-understood convex optimization methods which may be readily invoked for large-scale antenna arrays. Furthermore, the performance of the nML detector is superior to that of linear detectors both in terms of its MSE and bit error rate (BER). Compared to the multiuser message passing based detector of \cite{wang2015multiuser}, clear benefits of the nML detector are that it is capable of detecting arbitrary constellations and it is robust to errors of channel estimation . {However, the number of iterations for nML detector to converge is around 20-40, which is still too large for practical mmWave massive MIMO systems.}

{A potential direction for future research is to extend the ML detector both to frequency-selective mmWave channels and to sparse mmWave channels.} Moreover, the low-complexity MMSE detector using convex optimization may be investigated to reduce the computational complexity at the cost of an affordable error-rate performance degradation.

\section{Other Key Aspects of Low-Resolution ADCs}
There are some other important challenges for the deployment of mmWave massive MIMO systems relying on low-resolution ADCs. In the following, we will discuss the issues of channel feedback, transmit precoding, and mixed-ADC architectures, respectively.

\subsection{Channel Feedback}
The channel feedback designed for quantized mmWave massive MIMO systems imposes the following challenges. Firstly, the phase information of the mmWave channel, which has to be fed back, cannot be accurately represented by low-resolution ADCs, whilst phase-invariant beamforming codebooks are not suitable. Secondly, {the channel norm becomes inaccurate because of} the sub-optimal threshold-setting of low-resolution ADCs. Both the angle of arrival (AoA) and the residual phase information should be quantized and {sent back} to the transmitter.

Therefore, a new feedback codebook designed for channel information feedback has to explicitly incorporate the quantized phase information of mmWave massive MIMO systems using low-resolution ADCs, where some feedback bits are used for conveying the AoA information, while the phase information of the channel is quantized by using the remaining bits. Moreover, the feedback delay and the channel errors imposed on it constitute key considerations in finite-rate feedback aided systems. {The channel impulse response feedback scheme (including the codebook design) exploiting the {sparse property} of mmWave channels is a promising topic for future work.} Additionally, joint channel estimation and feedback schemes should be considered, where the quantized output of ADCs decide the codebook index.


\subsection{Transmit Precoding}

With aid of having an estimation of the CSI to be encountered during the next transmission, a substantial performance improvement can be provided for realistic quantized MIMO systems by employing transmit precoding. The transmit precoding design of realistic quantized MIMO systems requires that the phase information of the channel matrix is quantized, while the amplitude of each matrix element is fixed \cite{jacobsson2016massive}.

The related research attempted to tackle this challenge. As discussed in \cite{mo2015capacity}, low-complexity transmit precoding relying on channel inversion is capable of approaching the capacity, {if the MIMO channel matrix has a low condition number and full row-rank}. {However, this assumption has a finite applicability for the highly correlated mmWave channels of small cells, since it becomes a challenge to distinguish the signal transmitted to the different users. Hence, the optimal design of the transmit precoding matrix for mmWave channels is scenario-dependent.} Continued effort is needed for developing an efficient transmit precoding scheme for mmWave massive MIMO systems {with ADC quantizers}, especially for multiuser scenarios.

\subsection{Mixed-ADC Resolution}

{Due to the aforementioned concerns of coarse quantization, many research has focused on the mixed-ADC massive MIMO system concept \cite{liang2015mixed,zhang2017mixed}. {By using a few of high-resolution ADCs}, the mixed-ADC architecture is expected to become inherently immune to the capacity loss, to the huge overhead of channel estimation, and to the error floor of symbol detection.} The quantization noise of channel estimation is reduced with the employment of high-resolution ADCs, t. Furthermore, the pilot overhead may also be reduced. Therefore, the channel of all antennas can be accurately estimated with the assistance of high-resolution ADCs. In the following signal detection process, the high- and low-resolution ADCs are connected to different receive antennas.


Given its compelling benefits, an efficient signal detector using this mixed-ADC receiver architecture should be developed for reducing the computational complexity imposed. Additionally, the optimal ADC assignment scheme associated with specific SNRs remains an open research problem. For example, low-resolution ADCs become less beneficial at high SNRs. A natural question is in terms of energy efficiency, how many high-resolution ADCs should be used at low, medium and high SNRs, respectively? Finally, {the performance analysis and signal detector design of mixed-ADC based massive MIMO systems communicating over wideband frequency-selective mmWave fading channels require further research efforts.}

\begin{figure}[t]
\centering
\includegraphics[scale=0.65]{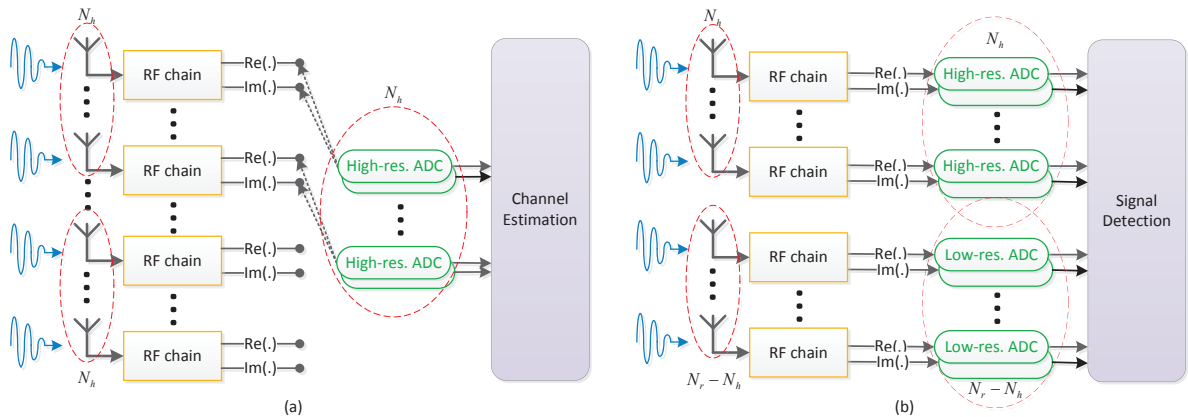}
\caption{The mixed-ADC architecture: a) Channel estimation using high-resolution ADCs: at different time slots, $N_h$ high-resolution ADCs are connected to different groups of $N_h$ receive antennas. The good-quality CSI of all antennas can be derived in a round-robin manner. b) Signal detection using high- and low-resolution ADCs: $N_h$ antennas are connected with $N_h$ high-resolution ADCs, the rest of antennas are employed with $N_r-N_h$ low-resolution ADCs. The AGC and VGA components are omitted here for simplified.
\label{fig:mixed_ADC}}
\end{figure}


\section{Conclusions}\label{se:conclusion}
{Low-resolution ADCs exhibits some compelling benefits in terms of their reduced hardware cost and power consumption in the 5G mmWave massive MIMO systems. However, the direct extension of information-theoretical analysis is unavailable of low-resolution ADCs. We detailed the challenges of realizing mmWave massive MIMO systems relying on low-resolution ADCs in this article. More specifically, we surveyed the relevant {challenges} of the quantized mmWave massive MIMO system model. Secondly, we discussed a variety of key physical-layer signal processing techniques, which are capable of enhancing the performance of mmWave massive MIMO systems relying on low-resolution ADCs. Finally, since THz frequencies exhibit similar channel characteristics to mmWave frequencies \cite{lin2015adaptive}, the methods discussed in this paper are also capable of circumventing some of the {issues} of practical THz massive MIMO systems.}
\bibliographystyle{IEEEtran}
\bibliography{IEEEabrv,1_bit_Ref}

\end{document}